\documentclass[vecphys]{svmult}

\usepackage{makeidx}         
\usepackage{graphicx}        
\usepackage{multicol}        
\usepackage{cite}            
\usepackage[bottom]{footmisc}

\makeindex             

\begin{document}

\title{Surface superconductivity controlled by electric field
}
\index{surface superconductivity}
\author{Pavel Lipavsk{\'y}$^{1,2}$ \and Jan Kol{\'a}{\v c}ek$^2$ \and 
Klaus Morawetz$^{3,4}$}
\institute{Institute of Physics, Academy of Sciences, 
Cukrovarnick\'a 10, 16253 Prague 6, Czech Republic
\texttt{kolacek@fzu.cz}
\and 
Faculty of Mathematics and Physics, Charles University, 
Ke Karlovu 3, 12116 Prague 2, Czech Republic
\texttt{lipavsky@fzu.cz}
\and University of Applied Science M\"unster,
Stegerwaldstrasse 39, 48565 Steinfurt, Germany
\texttt{morawetz@fh-muenster.de}
\and International Center for Condensed Matter Physics, 
Universidade de Bras\'ilia, 
70904-910, Bras\'ilia-DF, Brazil
}

\maketitle

\begin{abstract}
We discuss an effect of the electrostatic field on superconductivity 
near the surface. First, we use the microscopic theory of de~Gennes 
to show that the electric field changes the boundary condition for 
the Ginzburg-Landau function. Second, the effect of the electric field 
is evaluated in the vicinity of $H_{\rm c3}$, where the boundary 
condition plays a crucial role. We predict that the field effect on 
the surface superconductivity leads to a discontinuity of the 
magnetocapacitance. We estimate that the predicted discontinuity 
is accessible for nowadays experimental tools and materials.
It is shown that the magnitude of this discontinuity can be used to
predict the dependence of the critical temperature on 
the charge carrier density which can be tailored by doping. 
\end{abstract}

\noindent
\section{Introduction}
The surface of a superconductor is an important region in which the 
superconductivity nucleates and which represents a natural barrier
for penetrating or escaping vortices. It is desirable to control
surface properties so that the nucleation can be stimulated or 
suppressed. An even more attractive task is to open or shut the 
penetration barrier for vortices. A promising tool of the surface 
control is the gate voltage for
which we can benefit from the extensive technological experience 
with field effect transistors. 

Unfortunately, the interaction of the electric field applied to 
the metal surface with the superconducting condensate is very weak.
Indeed, the superconducting condensate does not interact with the 
electrostatic potential as shown by Anderson \cite{Anderson59}. 
The condensate feels only the indirect effects like changes of the 
local density of states or eventual changes of the surface crystal 
structure.

It is very likely that  it will become
possible to enhance the field effect on the superconductivity 
by a proper surface treatment.
To this end it would be of great advantage to understand how
the field interacts with the condensate and to have reliable
experimental methods directly aiming to measure the strength of this 
interaction. 

To support the experimental effort in this direction, in this 
chapter we provide a phenomenological theory of 
Ginzburg-Landau (GL) type supplemented with the de~Gennes 
boundary condition derived from the microscopic 
Bardeen, Copper, and Schrieffer (BCS) theory. 
It will be shown that the boundary condition captures the field 
effect on the condensate while the GL equation determines how 
the condensate responds to the field-affected boundary condition. 

The field effect on the superconductivity has been measured
under various conditions, nevertheless its actual strength is 
not yet accurately established. The most pronounced field 
effects are observed on thin layers, in which it is possible 
to increase or lower their critical temperature
\cite{GS60,XDWKLV92,FreyMannhart95,ATM03,MGT03}.
These samples are so thin that the applied field considerably 
changes the total density of electrons and the observed effect
can be interpreted in terms of the modified bulk properties. 
With thicker samples one meets the problem that the potential field
effect is restricted to the surface and the underlying bulk
overrides its contribution.

At the end of this chapter we discuss the field effect on the 
surface superconductivity near the third critical magnetic
field $H_{\rm c3}$. In this regime the bulk superconductivity
is absent and the surface superconductivity crucially depends
on the boundary condition. We will show that the field effect 
can be observed via the discontinuity in the magnetocapacitance
\cite{MLKB08,MLM08}.

\section{Limit of large Thomas-Fermi screening length}
\index{Thomas-Fermi screening}
To introduce the field effect on the superconductivity we start
from the theory employed by Shapiro and Burlachkov 
\cite{Shapiro84,Shapiro85,BKS93,Sh93,ShK93,SKB94} 
and by Chen and Yang \cite{Chen94}. It is justified for 
high-$T_{\rm c}$ superconductors in which the GL coherence 
length $\xi$ is very short, while the hole density is low leading
to relatively large Thomas-Fermi screening length $\lambda_{\rm TF}$.
In these materials $\lambda_{\rm TF}\sim\xi$, which allows us
to introduce field induced effects via local changes of the parameters 
of the GL theory.\index{Ginzburg-Landau theory}

Let us assume the jelly model \index{jelly model} in which the electric charge of
electrons is compensated by a smooth positively charged background.
Both charges are restricted to the half space $x>0$. 
The electric field applied to the surface is exponentially 
screened $E(x)=E\,{\rm e}^{-x/\lambda_{\rm TF}}$ inside the metal.
According to the Gauss equation $\epsilon\,{\rm div} E=\rho$,
the induced electron density $\delta n=\rho/e$ reads
\begin{equation}
\delta n(x)={\epsilon E\over e\lambda_{\rm TF}}
\,{\rm e}^{-x/\lambda_{\rm TF}}.
\label{a1}
\end{equation}

In the GL equation 
\begin{equation}
{1\over 2 m^*}\left(-i\hbar\nabla-e^*{\bf A}\right)^2\psi+
\alpha\psi+\beta|\psi|^2\psi=0,
\label{a2}
\end{equation}
the change of the electron density leads to changes of the GL
parameters $\alpha$ and $\beta$. \index{Ginzburg-Landau theory! parameters}
According to the Gor'kov theory\cite{Gor59} \index{Gor'kov theory}
$\beta={6\pi^2k_{\rm B}^2T_{\rm c}^2\over
7\zeta(3)E_{\rm F}n}$. It is thus a robust 
parameter in which $\delta n$ can create relative changes of
the order of $\delta\beta/\beta\sim \delta n/n$. These changes
can be neglected. The other parameter
$\alpha=\alpha'(T-T_{\rm c})$ is a difference of two large 
constituents. Here, $\alpha'={6\pi^2k_{\rm B}^2T_{\rm c}\over
7\zeta(3)E_{\rm F}}$ is modified also only negligibly as 
$\delta\alpha'/\alpha'\sim \delta n/n$. But, for temperatures 
close to the critical temperature, $T\to T_{\rm c}$, even small 
changes in $T_{\rm c}$ lead to large relative changes of $\alpha$ 
eventually even changing its sign. The GL equation with the dominant 
part of the field effect thus reads
\begin{equation}
{1\over 2 m^*}\left(-i\hbar\nabla-e^*{\bf A}\right)^2\tilde\psi+
\alpha\tilde\psi-\alpha'{\partial T_{\rm c}\over\partial n}
\delta n\tilde\psi+\beta|\tilde\psi|^2\tilde\psi=0.
\label{a3}
\end{equation}
We have denoted the field affected GL function as $\tilde\psi$.
Initially we shall use the customary GL condition
${\partial\over\partial x}\tilde\psi=0$ for this equation.

Let us assume now that this theory holds also for conventional 
superconductors where one always finds a sharp inequality
$\lambda_{\rm TF}\ll\xi$. Below we confirm the result obtained 
with this unjustified assumption by using the well justified 
microscopic approach of de~Gennes.

We split the GL function according to  $\tilde\psi=\psi+\delta\psi$, 
where $\delta\psi(x)=\delta\psi\,{\rm e}^{-x/\lambda_{\rm TF}}$ 
is the part of field induced perturbation which changes
on the short scale $\lambda_{\rm TF}$ and $\psi$ covers 
the rest changing on the larger scale $\xi$. Our aim is to 
establish approximative $\delta\psi$ and to eliminate it,
so that in the second step we will be left with the GL equation
for the slowly varying function $\psi$.

The short scale component has an enormously large space gradient 
which dominates its contribution to the GL equation,
\begin{equation}
-{\hbar^2\over 2 m^*}\nabla^2\delta\psi(x)-
\alpha'{\partial T_{\rm c}\over\partial n}\delta n(x)\psi(x)
\approx 0.
\label{a4}
\end{equation}
Since in this approximation the function $\delta\psi$ is nonzero 
only in the narrow layer $x\sim\lambda_{\rm TF}$, we can neglect 
the space dependence of $\psi$ and use its surface value. 
Performing derivatives one finds
\begin{equation}
\delta\psi=-
{2 m^*\lambda_{\rm TF}^2\over\hbar^2}
\,\alpha'{\partial T_{\rm c}\over\partial n}
{\epsilon E\over e\lambda_{\rm TF}}\psi(0).
\label{a5}
\end{equation}

Besides the very local contribution expressed by 
the function $\delta\psi$, the electric field induces also 
a perturbation on the scale of the GL coherence length $\xi$. 
Indeed, the GL boundary condition demands the zero derivative of 
the total function 
\begin{equation}
{\partial\over\partial x}(\psi+\delta\psi)=0
\label{a6}
\end{equation}
and the local part has a nonzero derivative 
\begin{equation}
{\partial\over\partial x}\delta\psi(x)|_{x=0}=-{\delta\psi\over
\lambda_{\rm TF}}.
\label{a7}
\end{equation}
From the GL boundary condition (\ref{a6}) and relations 
(\ref{a5}) and (\ref{a7}) one thus finds the boundary condition for 
$\psi$
\begin{equation}
\left.{\partial\psi\over\partial x}\right|_{x=0}=
{2 m^*\over\hbar^2}
\,\alpha'{\partial T_{\rm c}\over\partial n}
{\epsilon E\over e}\psi(0).
\label{a8}
\end{equation}
This relation can be interpreted as a field-affected GL boundary
condition. 

In usual problems handled by the GL theory one ignores the exact 
gap profile at the surface and focuses on its behavior deeper in 
the bulk on the scale $\xi$. In the same spirit we can ignore the 
short scale component $\delta\psi$ using the approximation $\tilde\psi
\approx\psi$. Doing so, we have to keep in mind that the short scale 
component leads to the field-affected boundary condition (\ref{a8}).
Solving the GL equation (\ref{a2}) with the boundary condition
(\ref{a8}) one obtains the GL function where the field effect manifests 
itself on the scale $\xi$.  

\section{de~Gennes approach to the boundary condition}
\index{de~Gennes boundary condition}
The limit of large screening length does not apply to conventional
superconductors. In fact, in metals the Thomas-Fermi screening is 
smaller than the interatomic distance and the jelly model is not 
justified to describe the interaction of the surface with the
electric field. Naturally, the gradient correction represented by
the kinetic energy of GL theory is not a sufficient approximation
of the non-local part of the BCS interaction kernel. 

To access short screening lengths, Shapiro and Burlachkov 
\cite{BKS93,Shapiro84,Shapiro85,Sh93,ShK93,SKB94} have employed 
a more sophisticated version of the GL theory in which the `kinetic 
energy' is not a mere parabolic function of the gradient but 
includes all derivatives up to infinite order in a form of the
di-gamma function. Since high order gradients are important only
in the short scale component, the above approach can be easily
modified in this way. One merely replaces the kinetic
energy in (\ref{a4}) by the corresponding di-gamma expression.

Although this high order gradient correction is elegant and simple,
it is likely not sufficient to cover changes on the sub-\AA ngstr\"om
scale, i.e., on the scale typical for majority of metals. Apparently,
one should employ the microscopic approach of Bogoliubov-de~Gennes
type pioneered by Koyama \cite{Koyama01} and other groups 
\cite{Machida2002443,Machida03,Machida2003659,zha:104508,zhu:014501,Zhu2005420,Zhu2006237,Jin04}. 
Beside microscopic details of the gap
near the surface, one has to take into account that
the simple exponential decay of the charge from a sharp surface is
not a very realistic model of metals, because electrons tunnel out of 
the metal. Realistic studies of the surface are problematic, however, 
even in the normal state, namely due to the nontrivial exchange-correlation 
interaction at the strong gradient of the electron density innate to all 
surfaces. 

Instead of improving the above approach, it is advantageous to
formulate the boundary condition directly from the microscopic 
theory of de~Gennes \cite{G66}. De~Gennes did not assume the electric 
field explicitly. His result, however, does not specify the forces
forming the surface so that the applied field can be included.

Gor'kov has shown that the BCS gap $\Delta$ and the GL function are 
proportional to each other, $\psi={\rm const}\times\Delta$. Using an 
extrapolation of the BCS gap from the bulk towards the surface, 
de~Gennes has arrived at the boundary condition of the form
\begin{equation}
{1\over\psi(0)}\left.{\partial\psi\over\partial x}\right|_{x=0}
=
{1\over\Delta_0}\left.{\partial\Delta\over\partial x}\right|_{x=0}
=
{1\over b},
\label{b1}
\end{equation}
where $b$ is called the extrapolation length. 

Within the BCS theory, the extrapolation length is given by the formula
\begin{equation}
{1\over b}={1\over\xi^2(0)}
{1\over N_0V}
\int\limits_{-\infty}^\infty dx{\Delta(x)\over\Delta_0}
\left[1-{N(x)\over N_0}\right]
\label{b2}
\end{equation}
derived by de~Gennes (Eq. (7-62) in Ref.~\cite{G66}). Here $N(x)$
is the local density of states at the Fermi level and $N_0$ is its 
bulk limit. The notation of the gap is different. While $\Delta(x)$
is the true local value of the BCS gap, $\Delta_0$ is the fake 
surface value after all short scale components have been removed, 
i.e., it is a value extrapolated from the near vicinity to the surface. 
Finally, $V$ is the BCS interaction and $\xi(0)$ is the GL coherence
length at `zero' temperature. In pure metals it is linked to the BCS 
coherence length $\xi_0$ as $\xi(0)=0.74\,\xi_0$.

De~Gennes estimated a typical 
value of $b\sim 1$~cm at metal surfaces in vacuum. This value
is very large on the scale of the GL coherence length, therefore 
this contribution is usually neglected. The approximation, 
$1/b\approx 0$, corresponds to the original GL condition 
\mbox{${\partial\over\partial x}\psi=0$}. 

Our aim is to include the effect of electric fields on the extrapolation 
length $b$. We denote as $b_0$ the extrapolation length in the
absence of the applied electric field and $\delta(1/b)=1/b-1/b_0$
reflects variation of the inverse length. 
The electrostatic potential corresponding to the 
electric field modifies the potential profile near the surface.
It results in a change of the density of states $\delta N(x)$. 
The density of states affects the gap function and creates its
deviation $\delta\Delta(x)$. In the linear approximation from
(\ref{b2}) we find the change of the inverse extrapolation length
as
\begin{equation}
\delta\left({1\over b}\right)={1\over\xi^2(0)}
{1\over N_0V}
\int\limits_{-\infty}^\infty dx\left\{{\delta\Delta(x)\over\Delta_0}
\left[1-{N(x)\over N_0}\right]-{\Delta(x)\over\Delta_0}
{\delta N(x)\over N_0}\right\}.
\label{b3}
\end{equation}

To estimate this change we recall the local density approximation 
in which the local density of states is a function of the local 
density, $N(x)=N[\rho(x)]$. Since the charge density $\delta\rho(x)$, 
 which screens the applied electric field, spreads over a layer of a few
{\AA}ngstr{\"o}ms near the surface, the perturbed density of state
is restricted to this very narrow layer, too. We can thus neglect 
the $x$-dependence of $\Delta(x)$ in the second term and write
\begin{equation}
\delta\left({1\over b}\right)=
{1\over\xi^2(0)}{1\over N_0V}
\left\{
\int\limits_{-\infty}^\infty dx{\delta\Delta(x)\over\Delta_0}
\left[1-{N(x)\over N_0}\right]
-{\Delta(0)\over\Delta_0}
{\delta N^{(2)}\over N_0}
\right\},
\label{b4}
\end{equation}
where 
\begin{equation}
\delta N^{(2)}=
\int\limits_{-\infty}^\infty dx~\delta N(x)
\label{b5}
\end{equation}
is the total change of the density of states per area. 

To estimate the second term we assume that the local density of 
states achieves the bulk value on the scale of a few {\AA}ngstr{\"o}ms.
Since the gap function changes on the scale of the BCS coherence length
which is much larger, we can expect that in the region of non-zero
function $1-{N(x)/N_0}$ the gap function keeps its shape. Assuming that
$\delta\Delta(x)\approx \Delta(x)\delta c$, the last $x$-integration
becomes identical to the integral in the de~Gennes condition (\ref{b2})
so that (\ref{b4}) simplifies to
\begin{equation}
\delta\left({1\over b}\right)=
{\delta c\over b}-{1\over\xi^2(0)}{1\over N_0V}{\Delta(0)\over\Delta_0}
{\delta N^{(2)}\over N_0}.
\label{b6}
\end{equation}

The relative change of the gap $\delta c$ can be estimated from 
the GL theory. The GL function obtained with the boundary condition 
of a large but finite extrapolation length $b$ can be written as a sum 
of the constant bulk term $\psi_\infty=\sqrt{-\alpha/\beta}$ 
and a small perturbation $\psi'$, i.e. $\psi=\psi_\infty+\psi'$. 
For the moment we ignore the magnetic field and assume a 
real GL function so that the GL equation reads 
$\xi^2\nabla^2\psi-\psi+\psi^3/\psi_\infty^2=0$.
Since $\psi'$ is proportional to $1/b$, we keep its linear
terms only, $\xi^2\nabla^2\psi'+2\psi'=0$. This equation has the
exponential solution $\psi'(x)=-\psi_\infty{\rm e}^{-\sqrt{2}x/\xi}
\xi/(\sqrt{2}b)$.
Since the GL coherence length $\xi$ is much smaller than the 
extrapolation length $b$, the exponential is small compared to
the constant term. Assuming that 
the GL function provides us with
the order of magnitude estimate of the gap function, we find that
$\delta c=\delta\Delta(0)/\Delta_0\approx\delta\psi(0)/\psi_0=-
(\xi/\sqrt{2})\delta(1/b)$. After substitution of this estimate into 
Eq.~(\ref{b6}) we obtain
\begin{equation}
\delta\left({1\over b}\right)=
-{1\over 1+{\xi\over \sqrt{2}b}}{\Delta(0)\over\Delta_0}
{1\over\xi^2(0)}{1\over N_0V}
{\delta N^{(2)}\over N_0}.
\label{b7}
\end{equation}

We can assume, that the induced density of states per area is
linearly proportional to the applied electric field,
\begin{equation}
\delta N^{(2)}=E\,g
\label{b8}
\end{equation}
so that the modified inverse length is of the form
\begin{equation}
\delta\left({1\over b}\right)={E\over U_{\rm s}},
\label{b9}
\end{equation}
where $U_{\rm s}$ has a dimension of a potential. According to
(\ref{b7}) this effective potential is given by
\begin{equation}
{1\over U_{\rm s}}=
-\eta\,
{1\over\xi^2(0)}{1\over N_0V}
{g\over N_0}.
\label{b10}
\end{equation}
We have introduced a dimensionless parameter
\begin{equation}
\eta={1\over 1+{\xi\over \sqrt{2}b}}{\Delta(0)\over\Delta_0},
\label{b11}
\end{equation}
which captures the effects of the gap profile near the surface.
According to de~Gennes' estimate \cite{G66}, the surface ratio 
$\eta$ is of the order of unity. 

Let us note, that a boundary condition similar to the de~Gennes boundary
condition described above can be derived also from the minimum free 
energy principle \cite{BoundCond09}.

\section{Link to the limit of large screening length}
To draw a link between the de~Gennes-type formula (\ref{b10}) and 
the field-affected GL boundary condition (\ref{a8}) obtained 
in the limit of large Thomas-Fermi screening lengths, we evaluate
the coefficients of the de~Gennes formula in the jelly model. As
above we assume that in the absence of the electric field the
extrapolation length diverges, $1/b_0=0$. 

For zero electric field the density of states is step-like, 
$N(x)=N_0$ for $x>0$ and $N(x)=0$ elsewhere. Now we include the 
electric field. It is exponentially screened due to the induced
electron density given by (\ref{a1}). We employ the local density 
approximation and assume that the local density of states is a 
function of the local density of electrons
\begin{equation}
N(x)=N_0+{\partial N_0\over\partial n}\delta n(x).
\label{c1}
\end{equation}
This approximation yields a simple change of the density of
states per area
\begin{equation}
\delta N^{(2)}={\partial N_0\over\partial n}
\int\limits_0^\infty dx\,\delta n(x).
\label{c2}
\end{equation}
The induced density of electrons per area is given by the
Gauss law
\begin{equation}
\epsilon E=-e\int\limits_0^\infty dx\,\delta n(x).
\label{c3}
\end{equation}
From relations (\ref{b8}), (\ref{c2}) and (\ref{c3}) we find the
coefficient of the density of states
\begin{equation}
g=-{\partial N_0\over\partial n}{\epsilon\over e}.
\label{c4}
\end{equation}

To be able to compare the BCS formula with the de~Gennes-type one,
we have to express both expressions in terms of the same parameters.
We thus convert the parameters of de~Gennes-type formula to their
phenomenological counterparts.

The GL coherence length $\xi=\hbar/\sqrt{-2m^*\alpha}$ depends on
the temperature according to $\xi=\xi(0)/\sqrt{1-T/T_{\rm c}}$, where $\xi(0)$
is the `zero' temperature value. From $\alpha=\alpha'(T-T_{\rm c})$
one finds 
\begin{equation}
{1\over \xi^2(0)}={2m^*\alpha'T_{\rm c}\over\hbar^2}.
\label{c5}
\end{equation}

Finally, we have to express the BCS interaction potential $V$ in
terms of the critical temperature. In the BCS critical temperature
\begin{equation}
T_{\rm c}=0.85\,\Theta_{\rm D}{\rm e}^{-{1\over N_0V}}
\label{c6}
\end{equation}
we assume that the Debye temperature $\Theta_{\rm D}$ and the BCS 
interaction $V$ are independent of the electron density. This
corresponds to approximations we have tacitly used above ignoring 
the electric field effect on the phonon spectrum. The density 
dependence of the critical temperature follows thus from the 
density dependence of the density of states
\begin{equation}
{\partial T_{\rm c}\over\partial n}=0.85\,\Theta_{\rm D}
{\rm e}^{-{1\over N_0V}}{1\over N_0^2V}
{\partial N_0\over\partial n}=T_{\rm c}{1\over N_0^2V}
{\partial N_0\over\partial n}.
\label{c7}
\end{equation}
We will use this relation to express density derivatives of
the density of states in terms of the density derivative of the
critical temperature.

Now we can rewrite the effective potential in terms of phenomenological
parameters. Using $\xi(0)$ from (\ref{c5}) in equation (\ref{b10})
we find
\begin{equation}
{1\over U_{\rm s}}=-\eta\,
{2m^*\alpha'\over\hbar^2}T_{\rm c}{1\over N_0^2V}g.
\label{c8}
\end{equation}
Next we substitute $g$ from (\ref{c4})
\begin{equation}
{1\over U_{\rm s}}=\eta\,
{2m^*\alpha'\over\hbar^2}T_{\rm c}{1\over N_0^2V}
{\partial N_0\over\partial n}{\epsilon\over e}.
\label{c9}
\end{equation}
The group of terms around $\partial N_0/\partial n$ can be substituted
with the help of (\ref{c7}) so that we obtain
\begin{equation}
{1\over U_{\rm s}}=\eta\,
{2m^*\alpha'\over\hbar^2}
{\partial T_{\rm c}\over\partial n}{\epsilon\over e}.
\label{c10}
\end{equation}

The boundary condition we have obtained now from the de~Gennes-type formula
\begin{equation}
\left.{\partial\psi\over\partial x}\right|_{x=0}
={E\over U_{\rm s}}\psi(0)=\eta\,
{2m^*\alpha'\over\hbar^2}
{\partial T_{\rm c}\over\partial n}{\epsilon E\over e}\psi(0)
\label{c11}
\end{equation}
differs from the large screening length limit (\ref{a8}) by the factor
$\eta$. Of course, a heuristic derivation of the field-effect from the 
GL equation cannot cover the factor $\eta$ which depends on the gap profile on
a scale smaller than the GL coherence length. 

In summary, the electric field applied to the surface of the superconductor
modifies the GL wave function near the surface. This effect is conveniently 
described by the GL theory, where the GL equation remains unaffected by the
field and the entire electric field effect is covered by a modified boundary
condition. In the next section we discuss an experiment which can be used to 
measure the predicted field effect on the GL boundary condition. 

\section{Electric field effect on surface superconductivity}
\index{surface superconductivity!electric field effect}
We will investigate now the magneto-capacitance 
for magnetic fields near the surface critical field $B_{\rm c3}$. 
We focus on this region since we expect that the bias voltage 
affecting only the surface has a relatively large effect on the 
surface superconductivity. In this section we show how  
the electric field affects the nucleation of superconductivity 
\cite{MLKB08}.

\subsection{Nucleation of surface superconductivity}
\index{surface superconductivity!nucleation}
At the surface critical field $B_{\rm c3}$ the superconductivity 
nucleates in the surface region. At the nucleation point the GL wave 
function $\psi$ is infinitely small, therefore we can work 
with the linearized GL equation, omitting the cubic term in
the equation (\ref{a2})
\begin{equation}
{1\over 2 m^*} (-i\hbar \nabla-e^*{\bf A})^2  \psi+\alpha  \psi=0.
\label{d1}
\end{equation}
The solution is restricted by the boundary condition (\ref{b1}).

The electrode is a superconductor which fills the half space
$x>0$. We assume a homogeneous applied magnetic field 
${\bf B}_{\rm a}=(0,0,B_{\rm a})$. Since an `infinitely' large 
electrode has translation invariance along the $y$ direction, 
we use the Landau gauge of the form 
\begin{equation}
{\bf A}=(0,B_{\rm a} x,0).
\label{d2}
\end{equation} 

Nucleation is possible if (\ref{d1}) has a nonzero solution,
i.e. if the parameter $-\alpha$ becomes 
equal to an eigenvalue $\varepsilon$ of the kinetic energy given 
by ${1\over 2 m^*}\left(-i\hbar\nabla-e^*{\bf A}\right)^2\psi=
\varepsilon\psi$. Since $\alpha$ changes with temperature,
$\alpha=\alpha'(T-T_{\rm c})$, the eigenvalue $\varepsilon$ of the 
kinetic energy determines the nucleation temperature $T^*$ according to 
$T^*-T_{\rm c}=-\varepsilon/\alpha'$. To avoid dual notation for
the same quantity, we will treat the equation as an eigenvalue for 
$\alpha$. Since $\alpha$ is negative, the nucleation temperature 
$T^*$ is always below the critical temperature $T_{\rm c}$ in the 
absence of the magnetic field. Note that we are looking for maximal
$\alpha$.

Assuming the translation invariance along the $y$ and $z$ axes we 
can write the wave function as
\begin{equation}
\psi(x,y,z)=\psi(x)~{\rm e}^{iky}{\rm e}^{iqz}.
\label{d3}
\end{equation}
Using (\ref{d3}) in the GL equation (\ref{d1}) we get a 
one-dimensional equation
\begin{equation}
{\hbar^2\over 2 m^*}\left(-\left({\partial\over\partial x}\right)^2+
\left(k-{e^*B_{\rm a}\over\hbar}x\right)^2+
q^2\right)\psi+
\alpha\psi=0.
\label{d4}
\end{equation}

Any non-zero value of $q$ results in the kinetic energy $q^2\hbar^2/2m$ 
which lowers the value of $\alpha$ reducing the nucleation temperature. 
The nucleation happens at the first possible occasion, i.e., at
the highest allowed temperature. We thus take $q=0$.
Similarly, we have to find the wave vector $k$ from the requirement 
of the highest nucleation temperature. 

\subsection{Solution in dimensionless notation}
It is advantageous to express the $x$-coordinate with the help of 
the dimensionless coordinate $\tau$
\begin{equation}
x=\tau l +2 l^2 k,
\label{coordinates}
\end{equation}
with the magnetic length 
\begin{equation}
l^2={\hbar \over 2 e^*B_{\rm a}}
\label{d6}
\end{equation} 
and the momentum $\tau_0=-2 k l$. The wave function is then 
proportional to the parabolic cylinder function of Whittaker 
\cite{MOS66}
\begin{equation}
\psi(x)={\cal N} D_{\tilde\nu}\left ({x\over l}+\tau_0\right ),
\label{d5}
\end{equation} 
which solves the differential equation (\ref{d1}) in the dimensionless 
notation
\begin{eqnarray}
{d^2 D_\nu(\tau)\over d\tau^2}&=&
\left ({\tau^2\over 4}-\nu-\frac 1 2\right )D_\nu(\tau) .
\label{d7}
\end{eqnarray}
The dimensionless boundary condition is
\begin{equation}
\left . {D_\nu'(\tau_0)\over D_\nu(\tau_0)}
\right |_{\tau_0=-2kl}
={\xi \over b} \sqrt{\nu+\frac 1 2},
\label{d8}
\end{equation}
where the prime denotes a derivative with respect to $\tau_0$.
The parameter 
\begin{equation}
\nu=-\frac 1 2 -{\alpha m^*\over e^* \hbar B_{\rm a}}
\label{d9}
\end{equation}
plays the role of an eigenenergy.

The boundary condition (\ref{d8}) links $\nu$ and $\tau_0$. 
Since we are looking for the minimal $\nu$, we take the
solution of (\ref{d8}) as a function $\nu(\tau_0)$. Besides 
the obvious numerical search we can give directly a nonlinear 
equation for this desired minimum given by $\nu'(\tau_0)=0$. 
For this purpose we differentiate (\ref{d8}) with respect to 
$\tau_0$  
arriving\footnote{With the 
minimum condition $\nu'=0$ at $\tau_0$, the derivative of 
(\ref{d8}) yields 
$$\left.{D^{\prime\prime}_{\bar\nu}\over D_{\bar\nu}}\right|_{\tau_0}=
\left.{D^{\prime 2}_{\bar\nu}\over D_{\bar\nu}^2}\right|_{\tau_0}.$$
The left hand side follows from (\ref{d7}) as 
$D^{\prime\prime}_{\bar\nu}/D_{\bar\nu}={\tau_0^2\over 4}-
\bar\nu-\frac 1 2$.
The right hand side is given by (\ref{d8}). The result is
$$\tau_0=\pm 2\sqrt{\left(1+{\xi^2\over b^2}\right)\left(
\bar\nu +\frac 1 2 \right)}$$
with the negative root being the physical one. Finally,
substituting this $\tau_0$ and a general relation
$D'_{\bar\nu}=\tau_0 D_{\bar\nu}/2-D_{\bar\nu+1}$ into 
(\ref{d8}) we obtain (\ref{d10a}).}
at \cite{MLKB08,MLM08}
\begin{equation}
\left . {D_{\tilde\nu+1}(\tau_0)\over D_{\tilde\nu}(\tau_0)} 
\right |_{\tau_0=-2 \sqrt{(\tilde\nu+\frac 1 2 )
(1+{\xi^2\over b^2})}}
\qquad=-\sqrt{{\left (\tilde\nu+\frac 1 2 \right )
\left (1+{\xi^2\over b^2}\right )}}-{\xi \over b}
\sqrt{\tilde\nu+\frac 1 2},
\label{d10a}
\end{equation}
where $\tilde\nu$ is the minimal value of $\nu$.

\begin{figure}[t]
\centering
\includegraphics*[width=.7\textwidth]{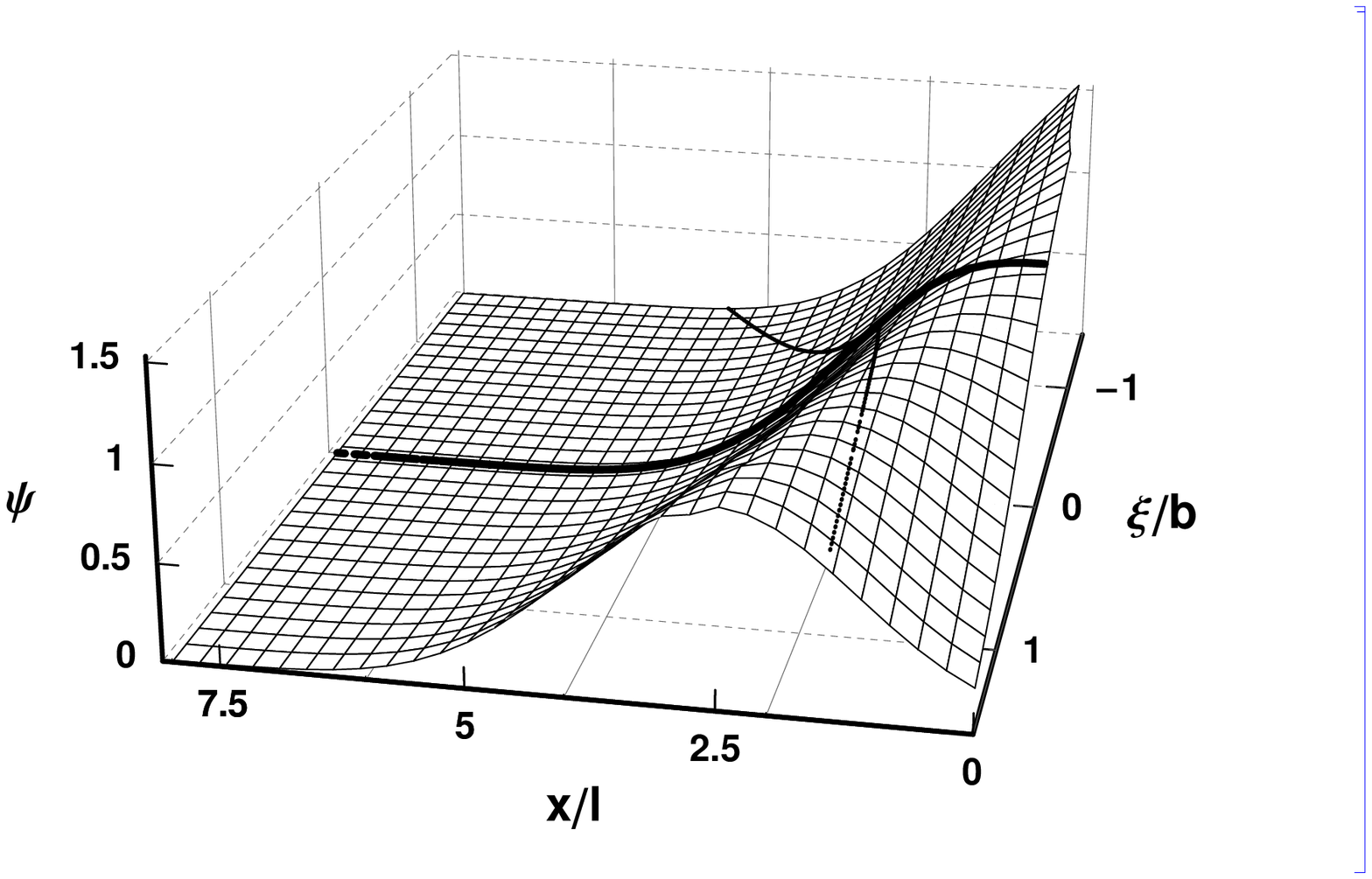}
\caption{The GL wave function of the condensate versus the 
boundary condition (\ref{b1}). The normal GL wave function 
without external bias is marked as thick line.}
\label{psi}
\end{figure}

Solving equation (\ref{d10a}) we find the minimal $\nu$ to
each given $\tau_0$, i.e., $\tilde \nu[\tau_0]$. Since
$\tau_0=-2 k l$, we find in this way the eigenenergy as a function 
of momentum $k$. 
The GL wave function (\ref{d5}) is now specified except for 
its amplitude ${\cal N}$. We discuss this amplitude below.
In figure~\ref{psi} we see how the shape of the GL wave function 
evolves with inverse extrapolation length $1/b$, i.e. 
how it depends on the external bias. For positive electric 
fields attracting charge carriers to the surface,
the superconducting density is pushed from the surface 
into the bulk while for oppositely directed electric fields 
the superconductivity is even more squeezed near the surface.
 
\begin{figure}[t]
\centering
\includegraphics*[width=.9\textwidth]{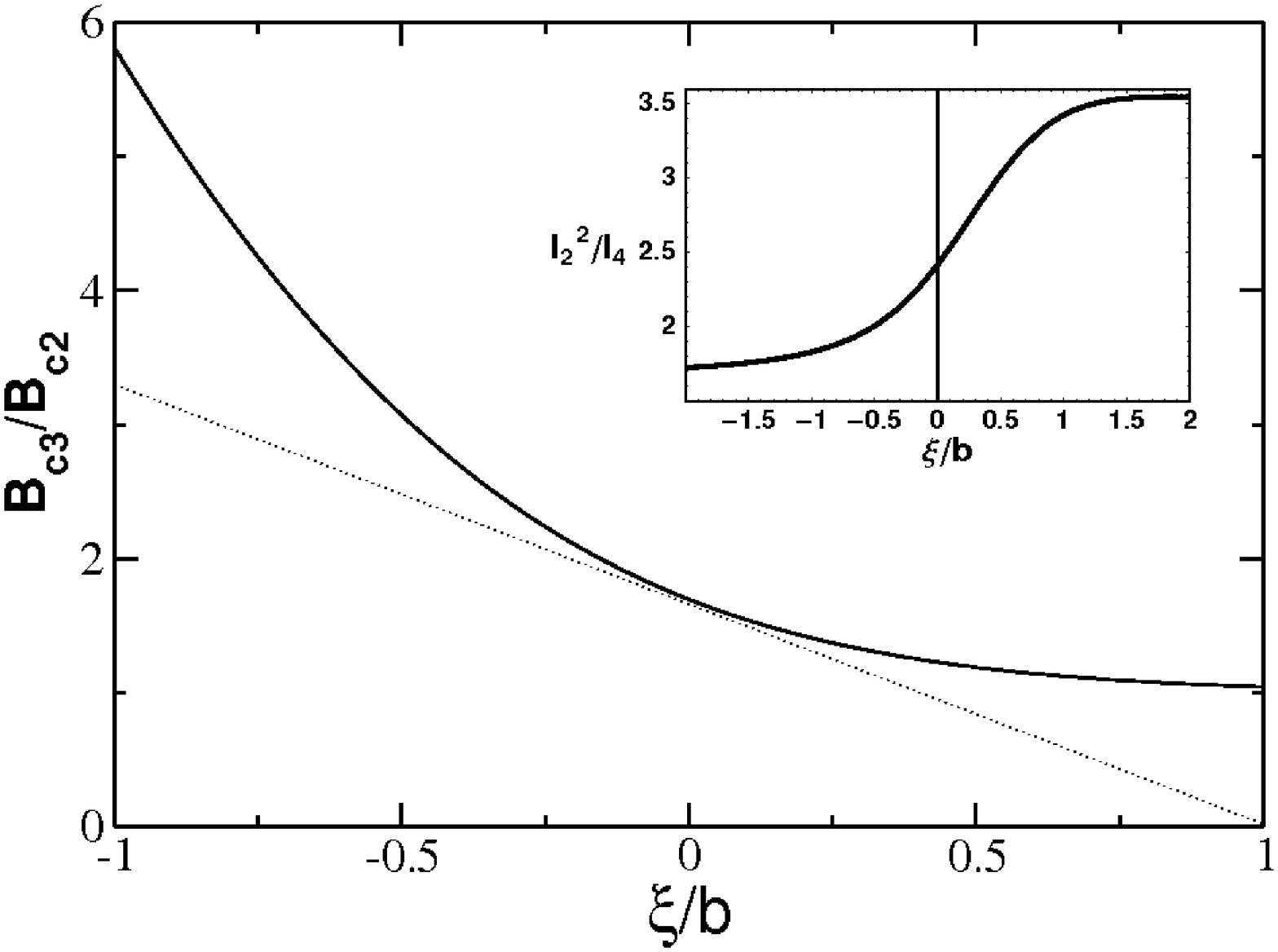}
\caption{The surface critical field $B_{\rm c3}$ versus 
extrapolation length. The exact solution (\ref{d10}) is 
the solid line and its slope at $\xi/b=0$ is shown
by the dotted line. The inset shows the ratio of integrals 
(\ref{d13}) and (\ref{d14}).}
\label{bc3f}
\end{figure}

The lowest eigenvalue $\tilde \nu$ corresponds to the highest 
attainable critical magnetic field
\begin{equation}
B_{\rm c3}={-m^* \tilde\alpha\over \hbar e^* (\tilde \nu+\frac 1 2)}
\equiv {B_{\rm c2} \over 2 \tilde \nu +1},
\label{d10}
\end{equation}
where $B_{\rm c2}$ is the upper critical field. 
In figure \ref{bc3f} we present the result for the surface 
critical field (\ref{d10}) versus the external bias. 
Without external bias the known GL solution \cite{SJdG63}
with $B_{\rm c3}/B_{\rm c2}=1.69461$ is reproduced.
We see that the external bias can enhance or decrease the 
surface critical value. 

\begin{figure}[t]
\centering
\includegraphics*[width=.9\textwidth]{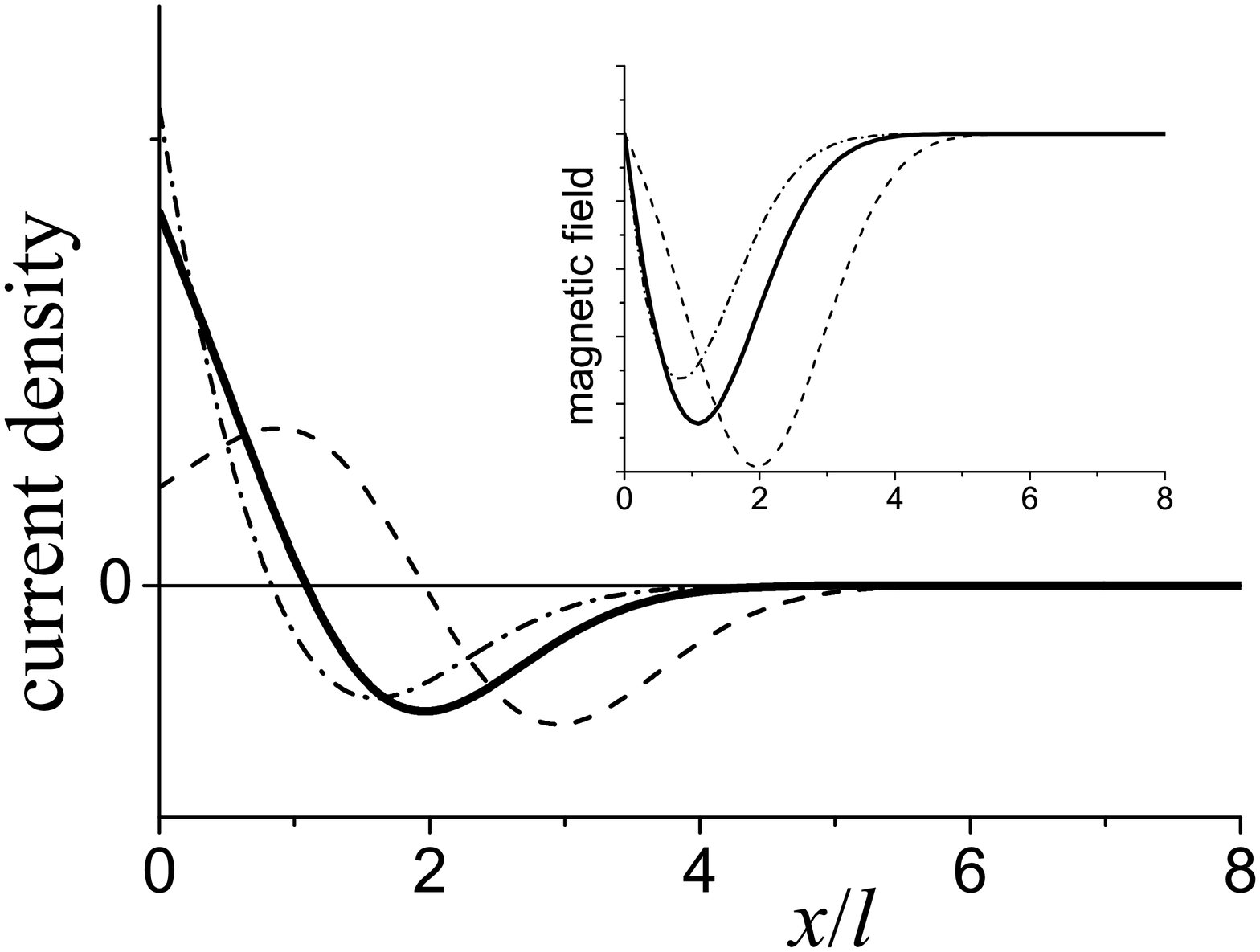}
\caption{Distribution of the current density for boundary conditions
with extrapolation lengths $\xi/b =$ 0, +1 and -1 
are plotted in bold, dashed and dash dot lines.
The inset shows the corresponding reduction of magnetic field.}
\label{fig_j}
\end{figure}

According to the GL wave function (\ref{d3}),
the current flows only in the $y$ direction. Its net value
given by the $x$ integral of the current density equals the 
$k$ derivative of the eigenenergy. Thus, for the minimal $\nu$ the
net current is zero. Current distributions for 
three different boundary conditions with extrapolation lengths  
$\xi/b =$ 0, +1 and -1 are plotted in the fig.(\ref{fig_j})
as bold, dashed and dash dot lines.     
Since there is no net current circulating around the sample, 
as is seen in the inset of the fig. (\ref{fig_j}), the magnetic field is reduced
only in the region of nucleation. Each surface thus acts independently. 

\subsection{Surface energy}

With the help of the GL wave function we can now calculate 
the surface energy
\begin{equation}
\sigma=
\int\limits_{0}^\infty d x \left 
[\alpha |\psi|^2+{\beta \over 2} |\psi|^4+
{[B_{\rm a}-B(x)]^2\over 2 \mu_0}+
{|(i\hbar {\bf \nabla}+e^* {\bf A})\psi|^2 \over 2 m^*}
\right ].
\label{d11}
\end{equation}
Here we neglect the free energy of the magnetic field, but
the term $\sim |\psi|^4$ which determines the amplitude of 
the GL wave function cannot be omitted even near 
the nucleation line. 

Since the applied field changes the shape of the GL wave 
function only weakly while its amplitude changes rapidly 
near the critical point, we fix the shape to be the nucleation 
function (\ref{d5}) and use the amplitude $\cal N$ as a 
variational parameter. Briefly we substitute $\psi={\cal N} 
D_{\tilde\nu}$ into (\ref{d11}) which gives
\begin{equation}
\sigma=(\alpha-\tilde\alpha){\cal N}^2 l\,I_2+
{1\over 2}\beta{\cal N}^4l\,I_4,
\label{d12}
\end{equation}
where $\tilde\alpha=-(\hbar e^*B/2m^*)\left(2\tilde\nu(E,B)+1\right)$ 
is the maximal eigenvalue corresponding to
the nucleation temperature under given magnetic and electric 
field, while $\alpha=\alpha'(T-T_{\rm c})$ is the GL parameter 
given by the actual sample temperature. The quadratic and quartic 
terms in (\ref{d12}) are weighted by dimensionless integrals
\begin{eqnarray}
I_2&=&\int\limits_{\tau_0}^\infty d \tau 
D_{\tilde\nu}^2(\tau)
,
\label{d13}\\
I_4&=&\int\limits_{\tau_0}^\infty d \tau  
D_{\tilde\nu}^4(\tau).
\label{d14}
\end{eqnarray}
In the normal state, $\alpha>\tilde\alpha$ and the minimum
of the surface energy is ${\cal N}^2=0$ giving $\sigma=0$. 
In the superconducting state $\alpha<\tilde\alpha$ and the 
minimum of $\sigma$ (\ref{d12}) is at ${\cal N}^2=-
(\alpha-\tilde\alpha)I_2/(\beta I_4)$ 
giving the surface energy
\begin{equation}
\sigma=-l {I_2^2\over I_4}{(\alpha-\tilde\alpha)^2\over 2\beta} .
\label{d15}
\end{equation}

At the critical point $\alpha=\tilde\alpha$ and the surface
energy vanishes. The surface energy and its first derivatives
with respect to the electric and magnetic field are continuous
at the critical point. The second derivatives are discontinuous.
In the next chapter we show how this discontinuity appears in the
magneto-capacitance.

\section{Magneto-capacitance}
We assume a capacitor in which the first electrode is a 
superconducting and the second electrode is a normal metal.
Our aim is to evaluate the contribution of the 
surface superconductivity to the capacitance.

The capacitance of the capacitor with one superconducting 
electrode reads
\begin{equation}
{1\over C_{\rm s}}={1\over C_{\rm n}}+
{1\over\epsilon^2S}
{\partial^2 \sigma\over\partial E^2},
\label{e1}
\end{equation}
where $S$ is the area of the capacitor, $C_{\rm n}$ is the 
capacitance when both electrodes are normal, $\epsilon$
is the permittivity of the ionic background in the superconductor. 
This follows 
from the inverse capacitance given by the second derivative
of the total energy $W$ with respect to the charge $Q$, 
$1/C={\partial^2W/\partial Q^2}$. The charge $Q$ is linked
to the electric field $E$ at the surface of the superconductor
via the Gauss law $\epsilon E=Q/S$.
Since energies of normal and superconducting states differ
by the surface energy $W_{\rm s}=W_{\rm n}+S\sigma$, one
arrives at equation (\ref{e1}) for the difference in capacitances.

\subsection{Discontinuity in magneto-capacitance}
Now we can evaluate the jump of the capacitance, which appears 
as the magnetic field $B$ exceeds the critical value $B_{\rm c3}$. 
Since $\alpha_E\to\alpha$ for $B\to B_{\rm c3}$, the discontinuity of
the inverse capacitance equals
\begin{equation}
{1\over C_{\rm s}}-{1\over C_{\rm n}}=-
{\hbar^2 e^{*2}B_{\rm c3}^2 l\,I_2^2\over
\epsilon^2m^{*2}S\beta I_4}
\left({\partial\tilde\nu\over\partial E}\right)^2,
\label{e2}
\end{equation}
where we have used ${\partial\alpha_E/\partial E}=-(\hbar e^*B/m^*)
(\partial\tilde\nu/\partial E)$. 

To evaluate the slope $\partial\tilde\nu/\partial E$ we recall the 
numerical result shown Fig.~\ref{bc3f}. The tangential dotted line 
yields $B_{\rm c3}/B_{\rm c2}=1/(2\tilde\nu+1)=
1.69-1.69\xi/b$ which follows also from an explicit variational calculation, see Eq. 20 in \cite{MLKB08}. Since $\partial(1/b)/\partial E=1/U_{\rm s}$, for
$1/b\to 0$ we find $(\partial\tilde\nu/\partial E)^2=
0.087\,\xi^2/U_{\rm s}^2$. From relation (\ref{c10}) thus follows
\begin{equation}
\left({\partial\tilde\nu\over\partial E}\right)^2=
0.087{4m^{*2}\epsilon^2\over\hbar^4 (e^*)^2}\xi^2\beta^2\eta^2
\left({\partial\ln T_{\rm c}\over\partial\ln n}\right)^2.
\label{e3}
\end{equation}
We have used the relation $\beta=\alpha'T_{\rm c}/n$ which holds for 
Gor'kov values of GL parameters. Substituting (\ref{e3})
into (\ref{e2}) we arrive at
\begin{equation}
{1\over C_{\rm s}}-{1\over C_{\rm n}}=-0.348\,{I_2^2\over I_4}
{B_{\rm c3}^2 l\over
\hbar^2S}
\xi^2\beta\eta^2
\left({\partial\ln T_{\rm c}\over\partial\ln n}\right)^2.
\label{e4}
\end{equation}

The GL coherence length $\xi=\hbar/\sqrt{2m^*\alpha}$ relates
to the surface critical field, see (\ref{d10})). For $1/b\to 0$
it yields $\xi^2=1.69\hbar/(e^*B_{\rm c3})$. Moreover, according to
(\ref{d6}) we can express $B_{\rm c3}$ via the magnetic
length, therefore 
\begin{equation}
{1\over C_{\rm s}}-{1\over C_{\rm n}}=-0.712\,
{1\over
(e^*)^2Sl}
\beta\eta^2
\left({\partial\ln T_{\rm c}\over\partial\ln n}\right)^2.
\label{e5}
\end{equation}
We have used ${I_2^2/ I_4}=2.42$, which is the value at $1/b=0$.

Since the GL parameter $\beta$ can be fitted from experimental 
results, relation (\ref{e5}) allows one to establish 
$\eta\,(\partial\ln T_{\rm c}/\partial\ln n)$. 
This material parameter describes the change of the critical temperature 
with the electron density.

\subsection{Estimates of magnitude}
For an estimate we assume some typical numbers. The most 
sensitive measurements of capacitance performed in the 
$C\sim\mu$F range are capable to monitor changes 
$\delta C/C\sim 10^{-6}$ with error bars at  $\delta C/C
\sim 10^{-7}$. From the capacitance $C=\epsilon_{\rm d}S/L$ 
one sees that a 1000\,\AA-thick dielectric layer with 
$\epsilon_{\rm d}=10^3\epsilon_0$ has an optimal area of 
$S=10$~mm$^2$ which is about the usual size of such samples 
\cite{Hwang02}.

To estimate $\beta$ we use Gor'kov's relation 
$\beta=6\pi^2k_{\rm B}^2T_{\rm c}^2/(7\zeta(3)E_{\rm F}n)$.
For Niobium $T_{\rm c}=9.5$\,K and $n=2.2\times 10^{28}$/m$^3$.
The free electron model used by Gor'kov than gives the Fermi
energy $E_{\rm F}=4.6~10^{-19}$~J. The corresponding GL parameters 
then is $\beta=1.2~10^{-53}$~Jm$^3$. The logarithmic derivative 
is estimated in \cite{LKMBY07} as
${\partial\ln T_{\rm c}/\partial\ln n}=0.74$. Since $\eta$ is
not known, we take $\eta=1$ according to the simple theory. 
Finally we need the third critical magnetic field $B_{\rm c3}$
\index{critical field $B_{\rm c3}$} 
to estimate the magnetic length $l$. From $B_{\rm c3}=1.69\,B_{\rm c2}$ 
and the experimental value $B_{\rm c2}=0.35$\,T \cite{FSS66} one 
finds $B_{\rm c3}=0.59$\,T, which yields $l=325$\,\AA. With 
these values from equation (\ref{e5})we obtain the discontinuity 
$1/C_{\rm s}-1/C_{\rm n}= -2.8~10^{-4}$~F$^{-1}$. Since the capacitance
was estimated to be $\sim 10^{-6}$~F, the corresponding relative change
$C_{\rm s}/C_{\rm n}-1\sim 3\times 10^{-10}$ is too small to be observed.
A slightly more optimistic estimation can be found in Ref.~\cite{MLM08}. 

In high-$T_{\rm c}$ materials the GL parameter $\beta$ is by three 
to four orders of magnitude larger than in conventional metals due to 
larger $T_{\rm c}$ and lower density of holes giving also lower Fermi 
energy. Moreover, the logarithmic derivative of the critical temperature 
is about ${\partial\ln T_{\rm c}/\partial\ln n}=-4.82$ as estimated 
in Ref.~\cite{LKMBY07}. Finally, higher critical surface
magnetic field $B_{c3}$ allows to reduce the magnetic length to values
limited rather by experimental facility. For a typical field of 
$10$\,T the magnetic length is $79$\,\AA. These factors together
provide an enhancement to values $1/C_{\rm s}-1/C_{\rm n}= 
-52$~F$^{-1}$ or $C_{\rm s}/C_{\rm n}-1\sim 5\times10^{-5}$
which is an experimentally accessible discontinuity.

\section{Summary}
We have shown that the electric field applied to the surface of the
superconductor modifies the boundary condition of the GL wave function.
Since the surface superconductivity is sensitive to this boundary
condition, we have discussed the influence of the electric field.
From the surface energy we predict that a planar capacitor with one 
normal electrode and the other electrode to be superconducting reveals 
a discontinuity of the capacitance at the third critical field 
$B_{\rm c3}$. This discontinuity is too small for capacitors from
conventional superconductors but it is large enough to be observed 
in capacitors with ferroelectric dielectric layers of a width of 
1000\,\AA\ and non-conventional superconductor electrodes.

\bibliographystyle{plain}
\bibliography{sem1,deform,bose,genn,sem2,solid,aa} 


\printindex
\end{document}